\documentclass[structabstract]{aa}
\usepackage{graphicx}
\usepackage{subfigure}
\usepackage{psfig}
\usepackage{natbib}
\usepackage[english]{babel}
\usepackage{url}
\newcommand{\rev}{}
\newcommand{\revb}{}

\begin{document}
   \title{MHD modeling of coronal loops: the transition region throat }

   \author{M. Guarrasi
          \inst{1,2}
          \and
          F. Reale\inst{1,3}
          \and
          S. Orlando\inst{3}
          \and
          A. Mignone\inst{4}
          \and
          J. A. Klimchuk\inst{5}
              }

   \institute{Dipartimento di Fisica \& Chimica, Universit\`a di Palermo,
              Piazza del Parlamento 1, 90134 Palermo, Italy
         \and
         now at: CINECA - Interuniversity consortium, via Magnanelli,
       6/3, 40033 Casalecchio di Reno (Bologna), Italy \\ e-mail: m.guarrasi@cineca.it
       \and
             INAF-Osservatorio Astronomico di Palermo, Piazza del Parlamento 1, 90134 Palermo, Italy
         \and
             Dipartimento di Fisica Generale, Universit\`a di Torino, via Pietro Giuria 1, 10125, Torino, Italy
             \and
             NASA Goddard Space Flight Center, Greenbelt, MD 20771, USA }

     \titlerunning{MHD modeling of coronal loops}

   \date{Received ; accepted }


  \abstract
   {The expansion of coronal loops in the transition region may considerably influence the diagnostics of the plasma emission measure. The cross sectional area of the loops is expected to depend on the temperature and pressure, and might be sensitive to the heating rate.}
   {The approach here is to study the area response to slow changes in the coronal
heating rate, and check the current interpretation in terms of steady heating models.}
   {We study the area response with a time-dependent 2D MHD loop model, including the description of the expanding magnetic field, coronal heating and losses by thermal conduction and radiation from optically thin plasma. We run a simulation for a loop 50 Mm long and quasi-statically heated to about 4 MK.}
   {We find that the area can change substantially with the quasi-steady heating rate, e.g. by $\sim 40$\% at 0.5 MK as the loop temperature varies between 1 and 4 MK, and, therefore, affects the interpretation of DEM(T) curves.}
   {}


   \maketitle

%
\section{Introduction}
\label{intro}

An important challenge in trying to understand the structure
and heating of solar active regions is to explain the relative
amounts of material at different temperatures. This is typically
expressed in terms of the differential emission measure
distribution, DEM($T$). Models that use steady heating and a flux
tube geometry that slowly expands with height are unable to
reproduce the observed distributions. However, realistic flux tubes
are expected to expand very rapidly with height between the
chromosphere and corona, where the plasma beta ($\beta = 8 \pi P /
B^2$) changes from being much greater than to much less than unity.
The thin transition region where temperatures increase from
chromospheric to coronal values occurs somewhere within this rapidly
expanding throat. Thus, we expect the cross sectional area of flux
tubes to be a strong function of temperature. If the dependence has
the right form, steady heating models may in fact be able to explain
the observed DEM(T). This was demonstrated by \cite{Warren_al_2010_b} for the core of an active region (the
generally hot central part of the active region that connects
opposite polarity moss regions, which are the transition region
footpoints of the hot loops).

A potential difficulty with this picture has been pointed out by \cite{Tripathi2010a}.
The height of
the transition region depends on the coronal pressure and so moves
up and down in response to changes in the coronal heating rate,
either fast or slow. If the heating rate increases, so too does the
pressure, and the transition region is forced downward. If the
heating rate decreases, some of the overlying pressure is relieved,
which allows the transition region to move upward. For gradual
heating variations, the amount of vertical displacement is given
approximately by

\begin{equation}
  \Delta z \approx -\frac{7}{6} H_g \ln \left(\frac{Q}{Q_0}\right) ,
  \label{eq:deltaz}
\end{equation}

\noindent where $H_g$ is the gravitational scale height of the
chromosphere (approximately 500 km), $Q$ is the new heating rate,
and $Q_0$ is the original heating rate. We have made use of the fact
that $Q \propto p^{7/6}$ for quasi-static evolution, where $p$ is the pressure. See \cite{Klimchuk_2006}
for a similar expression appropriate to
impulsive heating. A heating rate increase of a factor of 100, which
increases the loop apex temperature by a factor of 3.7, causes the
transition region to be displaced downward by approximately 2700 km.
This is not small compared to the length scale of expansion in the
flux tube throat \citep{Gabriel_1976,Athay1981a}.
Thus, the cross sectional area at any given
temperature $A(T)$ could change dramatically. This would alter DEM($T$), so
that a model is no longer compatible with observations. The very
sensitive dependence of the $A(T)$ relationship on the heating rate
suggests that too much fine tuning is required for a steady heating
model to have widespread applicability.

The last statement involves a big caveat, however. It assumes that
flux tubes remain rigid as the heating and pressure change. This is
not the case in a real MHD system. With increasing pressure, a flux
tube will expand laterally at the same time that the transition
region is pushed downward. Hence, the area at a given temperature
might remain constant, whereas it would decrease in a rigid tube. It
is possible that the $A(T)$ relationship in the transition region
could remain largely unchanged. (We note, however, that any lateral
expansion would be even more pronounced in the corona, so the
emission measure relationship between the transition region and
corona would then change.) The purpose of our study is
to determine how $A(T)$ varies in response to changes in the coronal
heating rate, especially slow changes corresponding to quasi-static
evolution. If there is a strong dependence on the heating rate, then
the steady heating interpretation of DEM($T$) should be reexamined.


In this paper we present a 2D-MHD loop model that naturally accounts for loop expansion  through the transition region.
We have used this model to study the plasma response to the heating into and around the moss regions. \rev{At variance with 1D models having a specified expanding cross section \citep{Emslie1992a,Mikic2013a}, we use a full MHD description, i.e., the full set of MHD equations is solved altogether in a 2D spatial domain, that allows us to describe a beta-changing system and to have feedbacks between plasma and magnetic field in the critical region, i.e. around the transition region. As a consequence, and at variance with previous works, in our model the loop cross-section area is also a function of time and changes as the heating changes. Our 2D cylindrical coordinates make the description realistic for a 3D structure with cylindrical symmetry, that is a good approximation for a coronal loop (but see \citealt{Malanushenko2013a} for deviations from circular loop cross-sections). We have focussed on the expansion around the transition region, not in the whole loop \citep{Peter2012a}. Our numerical experiment is devoted to answering a very specific question on the interaction between the coronal heating and the cross-section of the transition region. It is complementary to self-consistent 3D MHD descriptions of coronal boxes \citep{Bingert2013a}.}

In Sec.\ref{sec:model} we describe our model, the related equations and numerical code, the initial and boundary conditions, and how we set up the simulation of interest. In Sec.\ref{sec:simul} the simulation and its results are illustrated and discussed in Sec.\ref{sec:discuss}.

\section{The Model}
\label{sec:model}

The concept here is to study the reaction and readjustments of a magnetic loop and of the plasma confined therein under the effect of a gradual heating release inside it. We focus on a single loop and its surroundings, and therefore our approach is complementary to large scale modeling (e.g. \citealt{Carlsson_al_2010,Gudiksen2011a,Bingert_Peter_2011,Martinez-Sykora_al_2011_A,Martinez-Sykora_al_2011_B}).

We describe a coronal loop as a magnetic flux tube linking two far
locations of the solar chromosphere. The footpoints are so far from
each other that we can assume that they are rooted in independent
chromospheres. For simplicity of modeling, we then describe the loop
as straightened into a vertical flux tube linking two chromospheric
layers at opposite extremes of the geometric domain (top and bottom
boundaries). 
We keep the gravity
proper of a curved flux tube. The geometry of our domain is 2D
cylindrical ($r-z$).  The
spatial domain is much broader than the cross-section of a typical
loop. The proper loop corona forms as soon as we put a heating excess in
the region around the central axis ($r=0$) and not elsewhere. The
heating is transported along the magnetic field lines and makes
plasma expand from the chromospheres upwards, filling the space
between the footpoints in the central region. 

Our model considers the time-dependent MHD equations including gravity (for a curved loop), thermal conduction (including the effects of heat flux saturation), a heating function,
and radiative losses from optically thin plasma.

The MHD equations are solved in the non-dimensional conservative form:

\begin{equation}
  \label{MHD_mass_cons_eq_1}
  \frac{\partial \rho}{\partial t} + {\bf \nabla} \cdot \left( \rho {\bf u} \right) = 0
\end{equation}

\begin{equation}
  \label{MHD_mom_cons_eq_1}
  \frac{\partial \rho {\bf u}}{\partial t} + {\bf \nabla} \cdot \left( \rho {\bf u}  {\bf u} - {\bf B} {\bf B} + {\bf I} P_{t} \right) = \rho {\bf g}
\end{equation}

\begin{eqnarray}
  \label{MHD_Ene_cons_eq_1}
  \frac{\partial \rho E}{\partial t} + {\bf \nabla} \cdot \left[ {\bf u} \left( \rho E + P_{t} \right)  - {\bf B} \left( {\bf v} \cdot {\bf B} \right) \right] = \nonumber \\
  \rho {\bf u} \cdot {\bf g} - {\bf \nabla} \cdot {\bf F_{c}} - n_{e} n_{H} \Lambda\left( T\right) + H \left( r, z, t \right)
\end{eqnarray}

\begin{equation}
  \label{MHD_Induct_eq_1}
  \frac{\partial {\bf B}}{\partial t} + {\bf \nabla} \cdot \left( {\bf u}  {\bf B} - {\bf B} {\bf u}\right) = 0
\end{equation}

 \begin{equation}
 {\bf \nabla} \cdot {\bf B} = 0
 \label{Maxwell_02}
\end{equation}


where:

\begin{equation}
  \label{MHD_P_t}
  P_{t} = p + \frac{{\bf B} \cdot {\bf B}}{2}
\end{equation}

\begin{equation}
  \label{MHD_Ene_tot}
  E = \epsilon + \frac{{\bf u} \cdot {\bf u}}{2} + \frac{{\bf B} \cdot {\bf B}}{2 \rho}
\end{equation}

\begin{equation}
  \label{TC_flux_pes_summary}
  {\bf F_{c}} =  \frac{F_{sat}}{F_{sat} + \left| {\bf F_{class}} \right|} {\bf F_{class}}
\end{equation}

\begin{equation}
  \label{TC_flux_class_summary}
  {\bf F_{class}} = k_{||} {\bf \hat{ b}} \left( {\bf \hat{ b}} \cdot {\bf \nabla} T \right) + k_{\bot} \left[  {\bf \nabla} T - {\bf \hat{ b}} \left( {\bf \hat{ b}} \cdot {\bf \nabla} T \right) \right]
\end{equation}

\begin{equation}
  \label{TC_flux_abs_class_summary}
  \left| {\bf F_{class}} \right| = \sqrt{\left( {\bf \hat{ b}} \cdot {\bf \nabla} T\right)^{2}(k_{||}^{2} - k_{\bot}^{2}) + k_{\bot}^{2} {\bf \nabla} T^{2}}
\end{equation}

\begin{equation}
  \label{TC_flux_sat_summary}
 F_{sat} = 5\phi \rho c_{iso}^3
\end{equation}

\noindent are the total pressure, and total energy density (internal energy $\epsilon$, kinetic energy, and magnetic energy) respectively, $t$ is the time, $\rho = \mu m_{H} n_{H}$ is the mass density, $\mu = 1.265$ is the mean atomic mass (assuming metal abundance of solar values; \citealt{Anders_Grevesse_1989}), $m_{H}$ is the mass of hydrogen atom, $n_{H}$ is the hydrogen number density, $ {\bf u}$ is the plasma velocity, ${\bf g}$ is the gravity acceleration vector for a curved loop, ${\bf I}$ is the identity tensor, $T$ is the temperature, ${\bf F_{c}}$ is the thermal conductive flux (see Eq.~\ref{TC_flux_pes_summary}, ~ \ref{TC_flux_class_summary}, ~ \ref{TC_flux_abs_class_summary}, ~ \ref{TC_flux_sat_summary}), 
the subscripts $||$ and $\bot$ denote, respectively, the parallel and normal components to the magnetic field, 
$k_{||} = K_{||} T^{5/2}$ and $k_{\bot} = K_{\bot} \rho^2 / (B^2 T^{1/2})$ are the thermal conduction coefficients along and across the field, $K_{||}$ and $K_{\bot}$ are constants, 
$c_{iso}$ is the isothermal sound speed,
$\phi = 1$ is a free parameter,
and $F_{sat}$ is the maximum flux magnitude in the direction of ${\bf F_{c}}$.
$\Lambda\left( T\right)$ represents the optically thin radiative losses per unit emission measure derived from CHIANTI v.$~7.0$ database \citep{Chianti_I,Chianti_VI,Reale2012a} assuming coronal metal abundances \citep{Feldman_1992}, and $H \left( {\bf s}, t \right)$ is a  function of space and time describing the phenomenological heating rate (see Sect.\ref{sec:simul}). We use the ideal gas law, $P = (\gamma -1) \rho \epsilon$. We assume negligible viscosity and  resistivity, \rev{except for those intrinsic in the numerical scheme}.
%


The calculations are performed using the PLUTO code
\citep{Mignone_2007,Mignone_al_2011}, a modular, Godunov-type code
for astrophysical plasmas. The code provides a multiphysics, algorithmic modular environment particularly oriented toward the treatment of astrophysical flows in the presence of discontinuities as in the case treated here. The code is designed to make efficient use of massive parallel computers using the message-passing interface (MPI) library for interprocessor communications. The MHD equations are
solved using the MHD module available in PLUTO, configured to compute intercell fluxes with the Harten-Lax-Van Leer approximate Riemann solver, while second order in time is achieved using a Runge-Kutta scheme. A Van Leer limiter for the primitive variables is used. The evolution of the magnetic field is carried out adopting the constrained transport approach \citep{Balsara_Spicer_1999} that maintains the solenoidal condition $({\bf \nabla} \cdot {\bf B} = 0)$ at machine accuracy.

PLUTO includes optically thin radiative losses in a fractional step
formalism \citep{Mignone_2007}, which preserves the 2nd time
accuracy, since the advection and source steps are at least $2^{nd}$
order accurate; the radiative loss $\Lambda$ values are computed at
the temperature of interest using a table lookup/interpolation
method. The thermal conduction is treated separately from advection
terms through operator splitting. In particular we adopted the
super-time-stepping technique \citep{Alexiades_al_1996} which has
been proved to be very effective to speed up explicit time-stepping
schemes for parabolic problems. This approach is crucial when high
values of plasma temperature are reached (as during flares),
explicit scheme being subject to a rather restrictive stability
condition (i.e. $\Delta t \le (\Delta x )^{2} / 2 \eta $ where
$\eta$ is the maximum diffusion coefficient), as the thermal
conduction timescale $\tau_{cond}$ is shorter than the dynamical one
$\tau_{dyn}$ (e.g. \citealt{Orlando_al_2005_A,Orlando_al_2008}).

\subsection{The loop setup}
\label{sec:setup}



We address a typical active region loop, with half length $L = 3 \times 10^{9}$ ~ cm and  temperature  $T \sim 3 \times 10^{6}$ ~ K.

We choose as initial plasma condition a plane-parallel loop
atmosphere with the temperature and density profile from a model of
hydrostatic loop \citep{Serio_1981_bis} with an apex temperature
about $8 \times 10^{5}~K$. For the chromospheric part of the loop we
use a hydrostatic atmosphere with a uniform temperature (i.e. $10^{4}$ K).  The temperature spans from $10^{4}$ K
in the chromosphere to $8\times 10^{5}$ K in the corona. The density
spans from $\sim 10^{14}$~cm$^{-3}$ in the chromosphere to $\sim
10^{8}$~cm$^{-3}$ in the corona. This atmosphere is much more tenuous
and cool than the subsequently heated one. The corona and the chromosphere are connected through a thin transition region where the temperature jumps from $10^4$ K to $10^6$ K in less than $10^8$ cm.

\begin{figure}[!ht]               
\centering
  {\includegraphics[width=10cm]{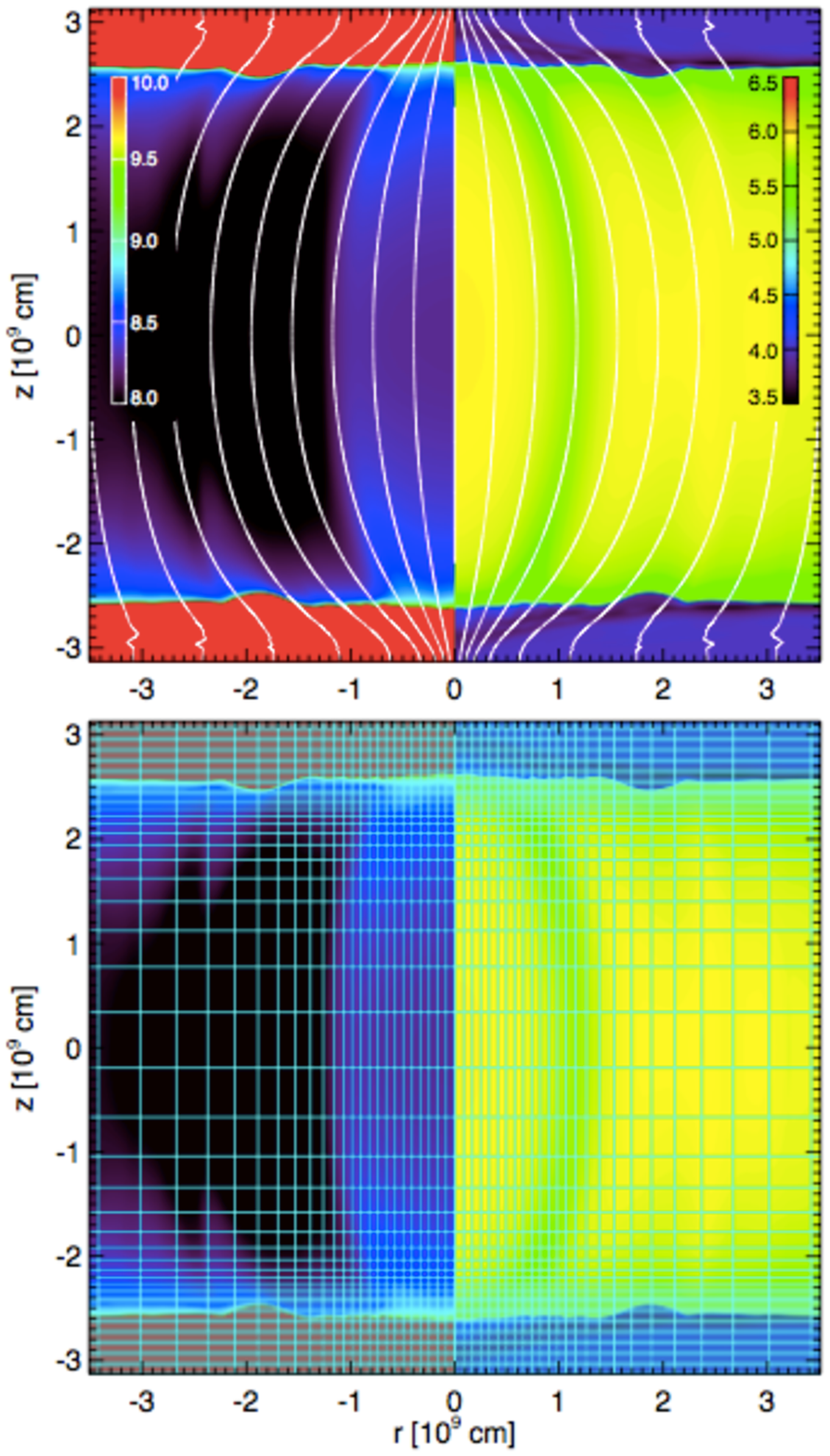}}
\caption{\small Top: Density (left, $[cm^{-3}]$) and temperature (right, $[K]$) color map (logarithmic scale) before the loop heating is switched on. The chromosphere is red in the density map and blue in the temperature map. The magnetic field lines are also marked (white lines). Bottom: pixels size map in the computational domain. Each box encloses 10$\times$10 grid points. }
\label{Fig:resol_init}
\end{figure}

The computational domain (see Fig.$~\ref{Fig:resol_init}$) extends from  $-3.1 \times 10^{9} ~ cm$ to $3.1 \times 10^{9} ~ cm$ in the $z$-direction (i.e along the loop axis),
and from  $r=0$ to $r = 3.5 \times 10^{9} ~ cm$ in the $r$-direction (i.e across the loop).

High resolution ($dr \sim dz \sim 3 \times 10^{6}$ cm) is needed
to describe appropriately the thin transition region from the chromosphere to the corona. 
The grid that we
use is non-uniform but fixed. Fig.~\ref{Fig:resol_init} shows the
pixel size map in the domain. We have the maximum resolution around
the transition region ($\left| z \right|
\approx 2.4 \times 10^{9} ~ cm$, layers where the colors change abruptly in Fig.~\ref{Fig:resol_init}) in the vertical direction and along
the loop axis in the $r$-direction ($r \approx 0$).

A fundamental ingredient of our study is that the loop where we
inject the heating expands from the chromosphere to the corona. We
want that the magnetic field has this expansion already at the
beginning of the heating release. Therefore, our strategy involves
two steps: first, we build the magnetic configuration in a tenuous,
static and plane parallel atmosphere, i.e. we build an ``empty''
loop. Then, we inject the heating in a localized region of the
domain to switch the loop on. For the first step, we start with a
magnetic field everywhere in the vertical direction, i.e. linking
the two dense independent chromospheres. {\it The magnetic field and the plasma density are more intense around the central axis, i.e. around $r=0$}. This
configuration is not at equilibrium, and a preliminary simulation is
devoted to let the configuration relax to equilibrium. The central initial density is chosen so that at equilibrium it is approximately equal to that of the surroundings.

In the tenuous corona, around the central axis the magnetic field
pressure is much higher than the plasma pressure, i.e. the ratio of
thermal pressure and magnetic pressure $\beta \ll 1$. As a result,
the magnetic field expands considerably in the corona. The magnetic
pressure is instead only a small perturbation in the chromosphere
($\beta \gg 1$) and it is left almost unchanged there. We let the
whole system evolve freely, keeping a small heating that compensates
for radiative and conductive losses. A quasi-steady state is reached after $t \sim 4000$ s (the Alfven and sound crossing time are $\sim
50$ s and $\sim 1000$ s, respectively). The result of this
preliminary step is a steady and stable loop atmosphere, with the
strong expansion of the magnetic field lines from the chromosphere
to the corona around the loop central axis, as we supposed
(Fig.~\ref{Fig:resol_init}). The magnetic field intensity decreases
from a few hundreds $G$ in the chromosphere to a few $G$ in the
corona (sufficient to confine the loop plasma). As shown in
Fig.~\ref{Fig:resol_init}, the area expansion is a factor of 6 from
the bottom of the chromosphere to the top of the
transition region, another factor 2 in the first 3000 km above the transition region, and another factor 2 to the top of the loop
(middle of the domain).

The plasma atmosphere also readjusts to a new equilibrium: in Fig.~\ref{Fig:resol_init} we see that in the magnetically-expanded configuration the atmosphere is no longer strictly plane parallel both in the temperature and in the density, but the inhomogeneities are small and leave the corona everywhere very tenuous and quite cooler than 1 MK. We also see some ripples in the chromosphere that do not affect the evolution when the loop heating is on. All plasma motions are a few km/s at most.

Many different initial conditions have been tested, finding that the resolution in the transition region is critical for accuracy.

We adopt axisymmetric boundary conditions at $r = 0$, i.e. along the symmetry axis, reflective boundary conditions at $r = 3.5 \times 10^9$ cm and reflective boundary conditions but with reverse sign for the magnetic field components at $z = \pm 3.1 \times 10^9$ cm.

\section{The loop simulation}
\label{sec:simul}

We use the results of the preliminary simulation as the starting point of a new simulation that studies the response of the loop to heating.

Our scope is to study what is the level of variability expected in
the moss at the base of a high pressure loop.  We check, for instance, whether the loop heating drives variations of the loop and in particular of its cross-section in the transition region. To this purpose, we consider the
situation in which the loop is heated as smoothly as possible (e.g.: \citealt{Winebarger_al_2011,Warren_al_2010_a,Warren_al_2010_b}). 


The heating function that we use is divided in two terms: the first
is the static term $H_{s}$ that covers uniformly the entire corona,
whereas the second $H_{t}$ is slowly varying
with time and covers only a part of the loop:

\begin{equation}
 H \left( r, z, t \right) = H_{s} + H_{t} \left(r, z, t    \right)
 \label{Eq:H_tot_mhd}
\end{equation}

According to the scaling law
\citep{Rosner_1978}, the static term is set to $H_{s} = 4.2 \times
10^{-5} ~ erg ~ cm^{-3}~ s^{-1}$, sufficient to sustain a static
loop with an apex temperature of about $8 \times 10^{5} ~ K$ (see
Section~\ref{sec:setup}).

The second term varies with time and space as:

\begin{equation}
 H_{t} \left({\bf s},t    \right) = H_{0} ~ f_r( r ) ~ f_z (z) ~ f_t(t)
 \label{Eq:H_time}
\end{equation}

where

\begin{equation}
  f_r \left(r    \right) = e^{-r^{2}/2\sigma_{H}^{2}}
 \label{Eq:H_space}
\end{equation}

\begin{equation}
  f_z(z) = \left\{ \begin{array}{ll}
       0 & z < -z_{0}   \\
      1 &  -z_{0} < z < z_{0} \\
         0 & z > z_{0}   \\
\end{array}     \right.
 \label{Eq:H_funct_fz}
\end{equation}

\begin{equation}
  f_t(t) = \left\{ \begin{array}{ll}
       0 & t < t_{0}   \\
      (t - t_0)/(t_1 - t_0) & t_0 < t < t_{1}
  \end{array}     \right.
 \label{Eq:H_funct_ft}
\end{equation}

\noindent $z_0 = 2.4 \times 10^9$ cm, $t_{0} = 0$, $t_{1} = 3600 ~ s$. The heating increases linearly in a time much longer than the typical dynamic and cooling times \citep{reale010}.  We set the maximum rate $H_{0} = 2 \times 10^{-3} ~ erg ~ cm^{-3} ~ s^{-1}$, that can sustain a loop with an apex temperature about $3 \times 10^{6} ~ K$ according to the scaling laws \citep{Rosner_1978}.
We have checked that the results do not change significantly if we assume a heating function that increases exponentially with a comparable rising time, \rev{and also if we consider a magnetic field with half intensity}.

The radial distribution of the heating has a Gaussian shape centered on the $r=0$ axis.  The width of the Gaussian is $\sigma_H = 3 \times 10^8$ cm, i.e. $1/10$ of the loop half-length. The heating distribution is uniform with height ($z$).

\subsection{Results}
\label{Sec:TaperResults}

\begin{figure}[!ht]               
\centering
  {\includegraphics[width=7cm]{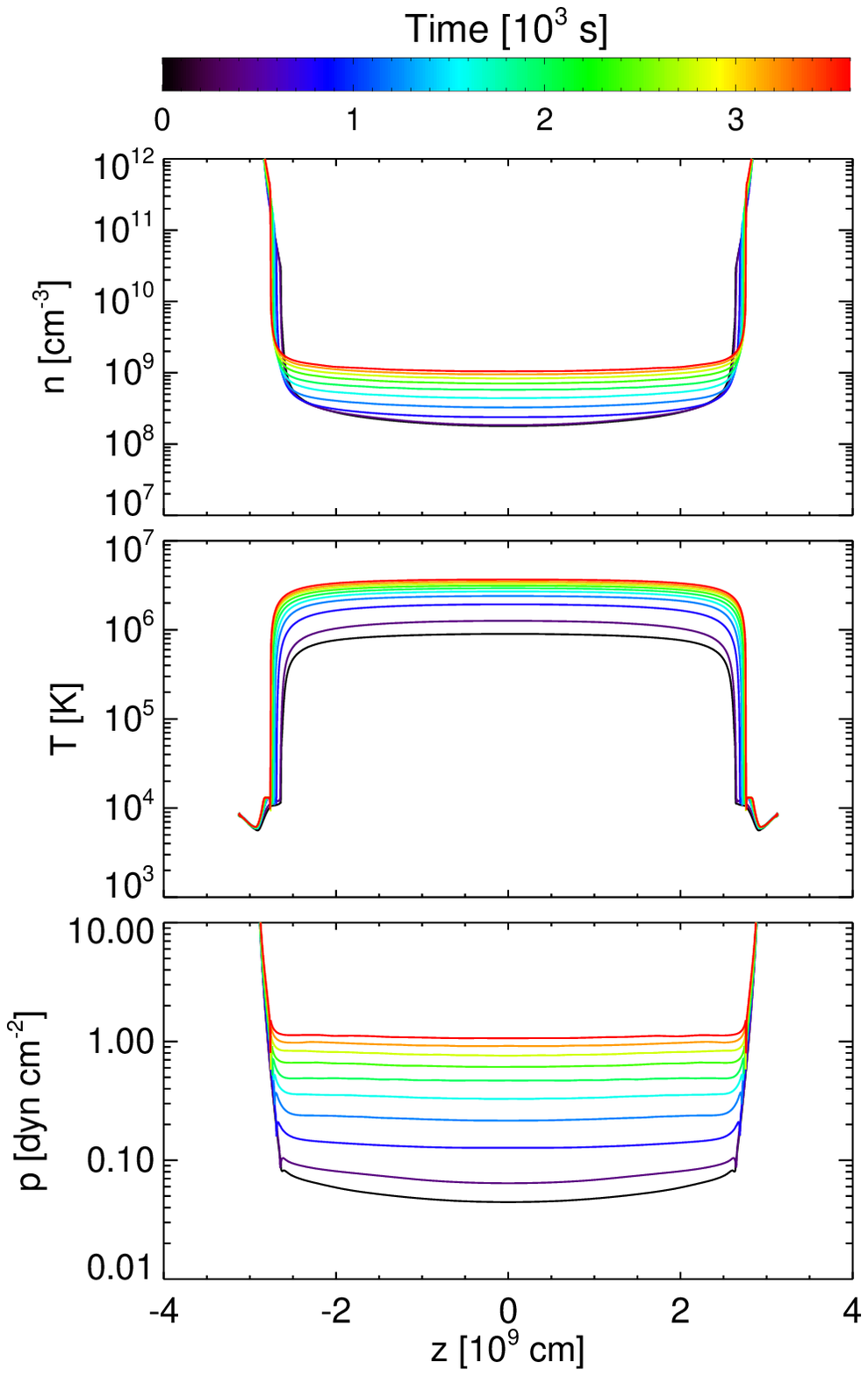}}
\caption{\small Evolution of the density, temperature and pressure profiles along the loop central axis ($r = 0$) in logarithmic scale after the time-dependent heating is switched on. Different line colors mark different times, from black (early) to red (late). The profiles are sampled regularly with a cadence of $\approx 400$ s. }
\label{Fig:timprof}
\end{figure}

The heating is released with gradually increasing intensity around
the loop central axis. Along this axis we expect an evolution very
similar to that obtained from a typical loop model. The profiles in
Fig.~\ref{Fig:timprof} show that this is indeed the case. The
temperature gradually increases in the heated region and reaches
$\sim 3$ MK when the heating gets to its maximum rate, i.e. $t \sim
3600$ s (red curves in Fig.~\ref{Fig:timprof}). As typical of
coronal loop models, the density increases -- gradually too,
due to the very slow growth rate of the heating -- while the
plasma "evaporates" from the chromosphere, due to the higher
pressure driven by the heating. The plasma dynamics
is minor at any time. The maximum apex density (at the final time) is $\sim 2 \times 10^9$
cm$^{-3}$.  The loop pressure increases uniformly in
the corona above 1 dyne/cm$^2$.

\begin{figure}[!ht]               
\centering
  {\includegraphics[width=10cm]{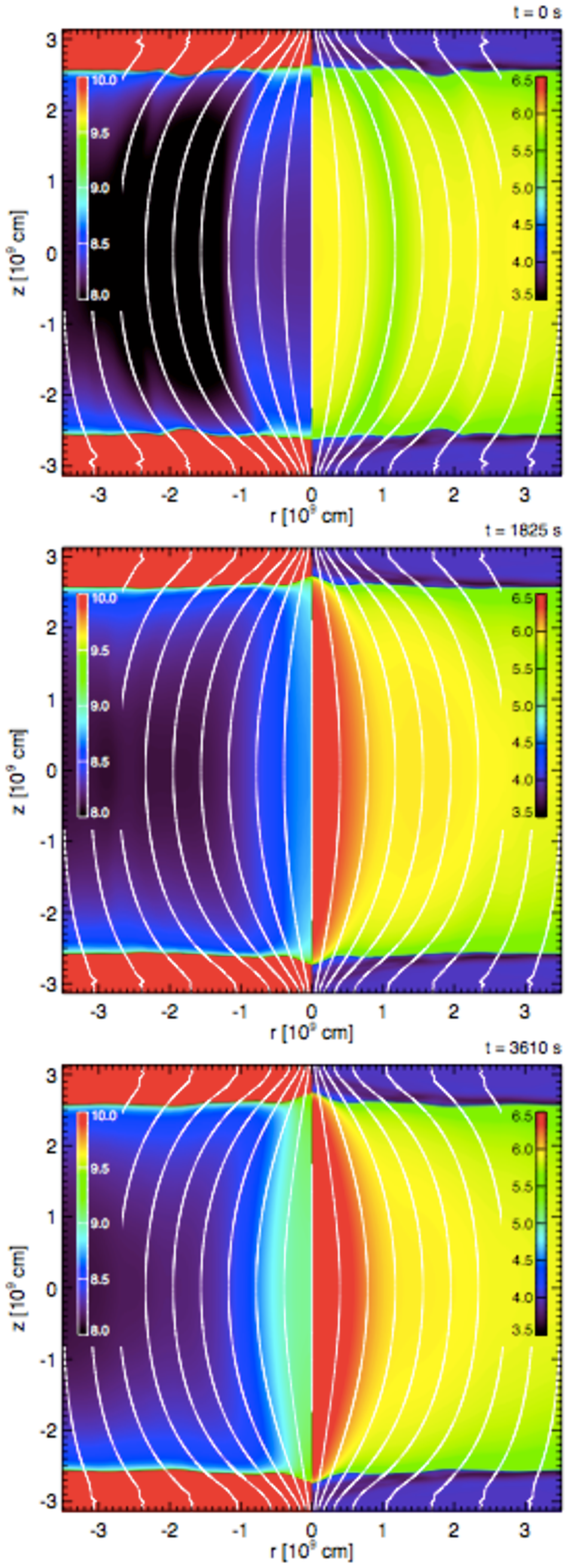}}
\caption{\small Density (left, $[cm^{-3}]$) and temperature (right, $[K]$) maps in logarithmic scale at three subsequent times after the loop heating is switched on: beginning (t = 0), half-way (t = 1825 s), and end of the heating (t = 3610 s). The magnetic field lines are also marked (white lines).  }
\label{Fig:timaps}
\end{figure}

Fig.~\ref{Fig:timaps} shows maps of temperature and density at three
different times, i.e. beginning, middle and end of the heating. The heating is conducted only along the magnetic
field lines, and the proper coronal loop forms and is clearly
visible at the end of the heating (bottom panel), both in the
temperature and in the density maps. No local features are visible\footnote{We have verified that, after the heating is switched off, the loop cools and drains back to conditions similar to the
initial ones.}. Overall, the evolution can be described as quasi-static.

\begin{figure*}[!ht]               
\centering
  {\includegraphics[width=14cm]{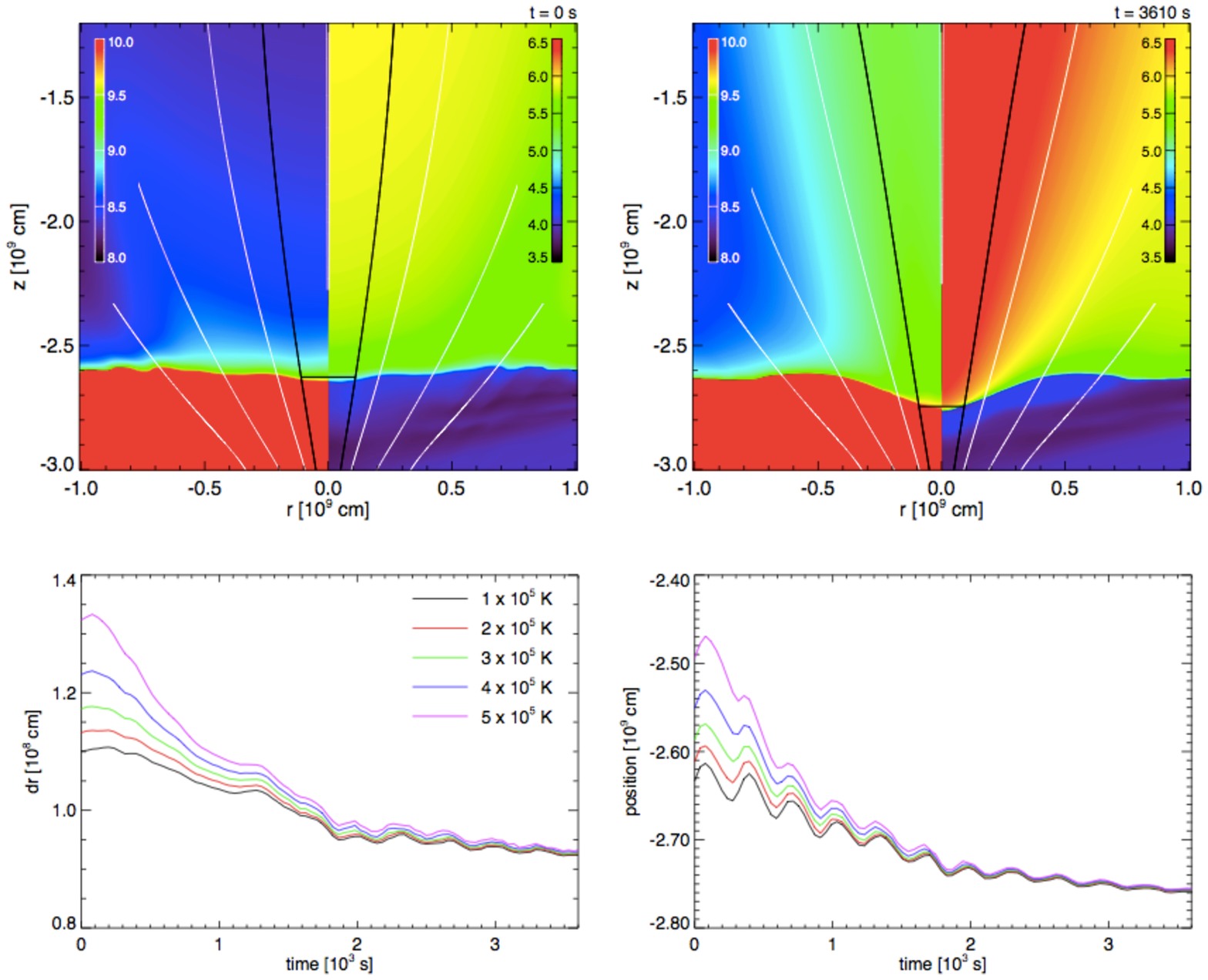}}
\caption{\small Evolution of the moss region and analysis of the transition region drift and area change: enlargement of density/temperature maps (Fig.~\ref{Fig:timaps}) in the transition region at two times t = 0 and t = 3610 s (see on-line Supplementary movie for an animated version of this evolution).  Magnetic field lines are marked (white and black lines). The black lines bound the region where we measure the cross-section. We also mark the relevant cross-section with a horizontal segment. The bottom panels show plots of the cross-section radius (left) and of the $z$ position of the layer at \rev{five linearly-increasing temperatures, i.e. 1, 2, 3, 4, $5 \times 10^5$ K} as a function of time, since the beginning of the time-dependent heating.}
\label{Fig:moss}
\end{figure*}

Our attention now turns more specifically to the transition region,
and in particular to the loop cross-section in the expansion region.
Fig.~\ref{Fig:moss} shows an enlargement around the region at one
loop footpoint at two different times. While at early times, before
the heating is switched on, the transition region layer is almost
flat with only small ripples, later we see that the transition
region of the heated part of the flux tube clearly drifts deeper in
the chromosphere, more where the heating is more intense, and forms
a kind of bowl. We now focus on layers at
certain fixed temperatures of the transition region, and check whether and to what extent the area of that section changes with time because of this
drift.

For quantitative analysis, we consider two magnetic field lines on
the sides of the heated part of the loop. 
\rev{The field lines are chosen to bound the width of the heated region in the corona (i.e. their distance corresponds to the $\sigma_H$ in the corona). 
These field lines are symmetric and separated from the central axis by $5 \times 10^7$ cm   at the lower boundary, $10^8$ cm in the transition region, and $4 \times 10^8$ cm at the loop apex (Fig.\ref{Fig:moss}). They enclose most of the heated region ($\sigma_H = 3 \times 10^8$ cm in Eq.\ref{Eq:H_space}; see Sect.\ref{sec:simul}).
We have checked that the results do not change much by choosing other field lines on both sides. The field lines experience some readjustments during the loop evolution, e.g. they become more straight in the low loop region (Fig.~\ref{Fig:moss}). This is because the pressure increase affects  the low-beta corona more than the chromosphere or transition region, causing it to expand to a greater degree.}

To study the moss visibility we concentrate on layers with fixed temperatures, that will be continuously visible in an EUV
narrow-band detector, for instance. We choose to measure the
position and width of the layer between the selected field lines at
temperatures \rev{increasing linearly from $10^5$ K to $5 \times 10^5$ K}
as the simulation progresses. Fig.~\ref{Fig:moss} shows that both
the width and the height decrease as the heating increases. As
mentioned above, the width varies considerably at all selected
temperatures by $\sim 40$\%, more at higher temperature. This occurs
because the layers deepen in the chromosphere by a few $10^8$ cm, in good agreement with the estimate from Eq.(\ref{eq:deltaz}).
The layer becomes progressively narrower with time, essentially
because the transition region
drifts deeper inside the chromosphere as the coronal pressure increases\footnote{As long as there are many
scale heights in the chromosphere, the pressure at the bottom of the
chromosphere, $P_{base}$, is practically independent of the 
coronal pressure. The top of the chromosphere (bottom of the
transition region) occurs at the height $z$ where $P = P_{corona} =
P_{base} \times \exp (-z/H_g)$.} in regions where $\beta$ gets smaller. This
variation is faster at the beginning and then slows down near the end of the simulation.  The minimum cross-section radius is $\approx 0.9 \times 10^8$ cm and is
nearly the same at all selected temperatures \rev{because the transition region gets thinner as the heating rate increases}.
Overall, since the area variation is a tracer of a variation of the
brightness, this result indicates that even slow variations of
heating lead to considerable variations of the moss appearance.

\rev{In both plots at the bottom of Fig.~\ref{Fig:moss}, we see periodic oscillations that modulate the decreasing trend. These are due to the heating process. 
The increasing heating drives moderate pressure fronts that hit and overshoot in the chromosphere. The period is approximately 300 s that is compatible with the traveling times of a perturbation at the sound speed ($c_s \sim 10^4 \sqrt{T} \sim 10^7$ cm/s) along a tube $5 \times 10^9$ cm long.}

\begin{figure}[!ht]               
\centering
 \centering
 \subfigure[]
  {\includegraphics[width=7.5cm]{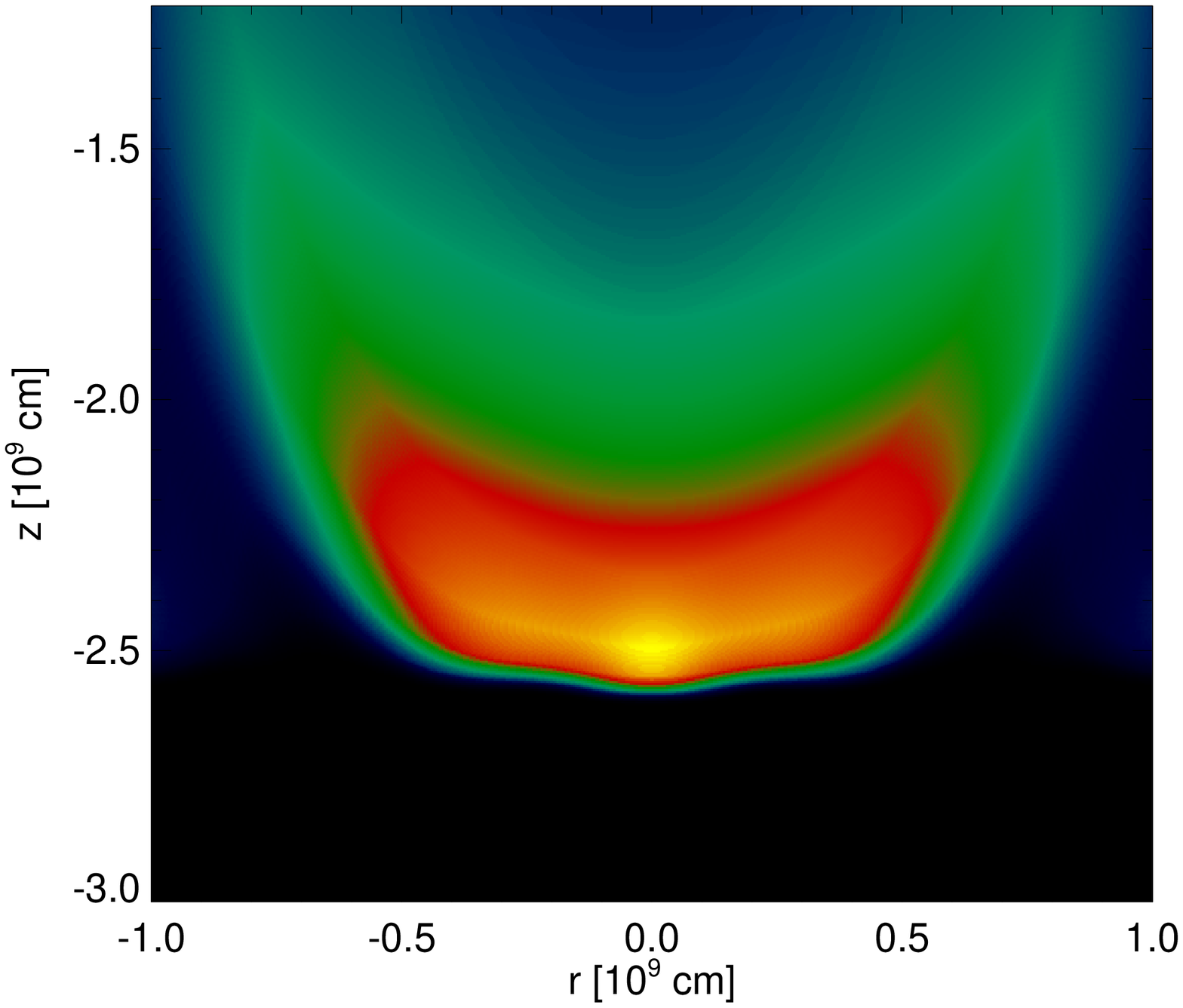}}
 \subfigure[]
  {\includegraphics[width=7cm]{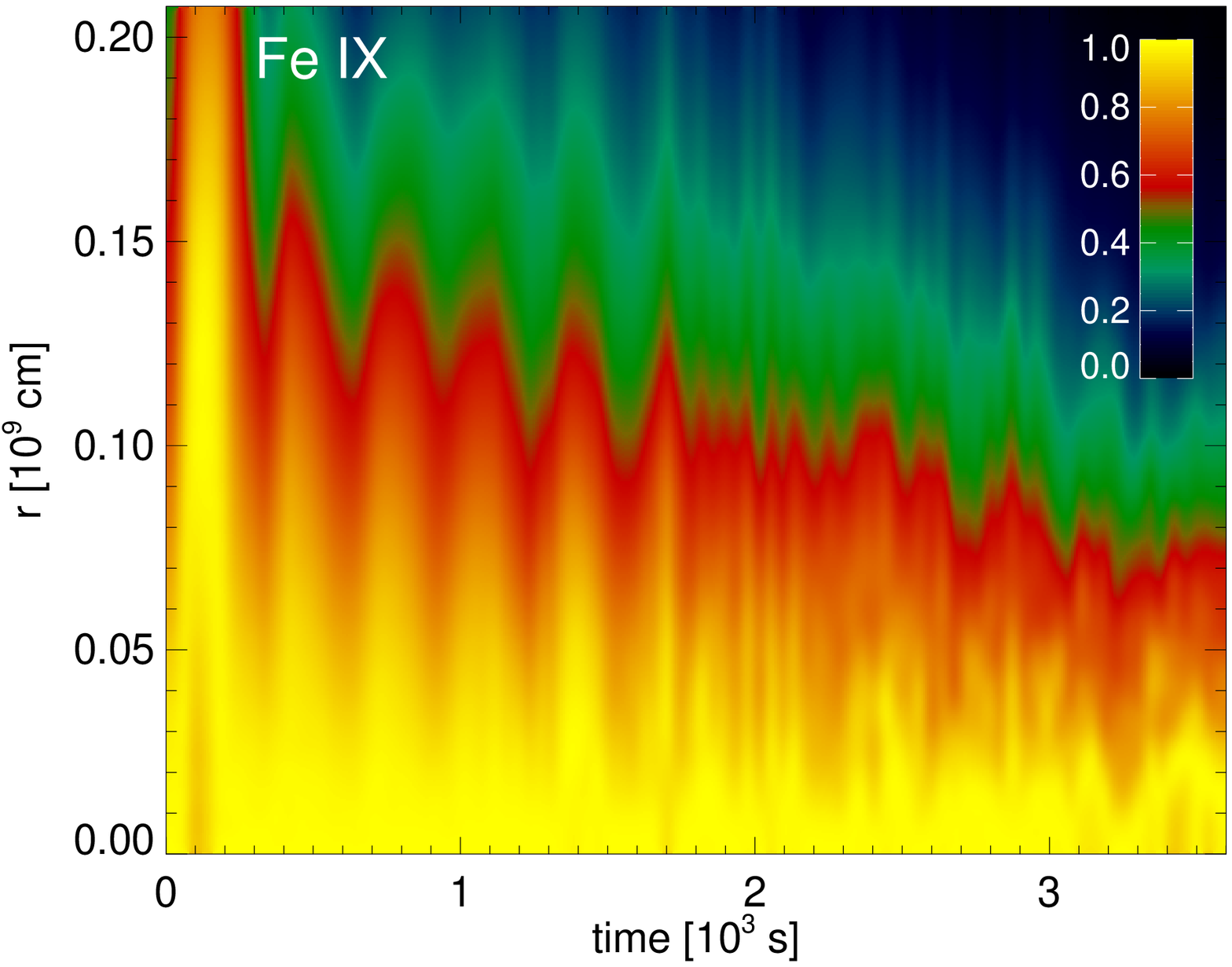}}
\caption{\small \revb{(a) Map of synthetic emission in the Fe IX 171 \AA\ line at t = 0, in the same enlargement as in Fig. 4. The emission is normalized to its maximum and the color scale is shown in panel (b).} (b) Time(x-axis)-space($r$, y-axis) plot of the synthetic emission in the Fe IX 171 \AA\ line. We plot the radial profile of the emission in the vertical direction at any time. Each profile is normalized to its maximum. }
\label{Fig:fe9}
\end{figure}

\rev{From the model results, we synthesize the Fe IX line emission. In 
particular, we derive the emission measure in the j-th domain cell as
em$_{j} = n_{Hj}^2 V_j$, where $n_{Hj}$ is the hydrogen number density
in the cell, and $V_j$ is the cell volume. From the values of emission
measure and temperature in the cell, we synthesize the corresponding
emission in the Fe IX line, using CHIANTI. \revb{Fig.~\ref{Fig:fe9}a shows the 2D map of the emission at time t = 0 in the same region as Fig.~\ref{Fig:moss}.} From the maps
at different times, we first derive the profiles of the emission along
the $r$-axis at each time $t$ by integrating the emission along the
$z$-axis. The time-space plot of the emission evolution is then derived from
these profiles, each normalized to its maximum. The result is shown in
Fig.~\ref{Fig:fe9}b. Looking, for instance, at the edge of the red color, we can clearly see that the emission shrinks as time progresses and the heating increases, coherently with the results shown in Fig.~\ref{Fig:moss}. The oscillations are emphasised because of the dependence on the square of the density. A factor 2 reduction of the radial size can be estimated.}

\section{Discussion}
\label{sec:discuss}

We investigated an MHD model of an active region loop, and in particular the variation of the loop cross-section in the moss region.

To study the low-lying parts of the loops in high detail, the assumption of constant cross-section all along the loop cannot hold, because it is known that the loop expands considerably going up from the transition region to the corona (e.g., \citealt{Gabriel_1976}). We could not help including this effect in the modeling, and how it changes in time and with the plasma $\beta$,  and this required a proper time-dependent magnetohydrodynamic description.

Our model considers the time-dependent MHD equations in a 2D cylindrical geometry including the gravity (for a curved loop), the thermal conduction, the coronal heating, and the radiative losses from optically thin plasma.

We first created the proper loop topology, with an expanding cross-section from the chromosphere to the corona, ready for the injection of heating. Then,  we heated our loop with a large-scale slowly-changing (quasi-steady) heating (e.g.,  \citealt{Winebarger_al_2011,Warren_al_2010_a,Warren_al_2010_b}). We focussed on the structure of the so-called moss in active regions, i.e. the thin loop layers at temperature ranging between about 0.1 and 1 MK, that are generally interpreted as the footpoints of hotter loops. We measured the position and size of the moss layers.

From Fig.\ref{Fig:moss} we have seen that the position and size of the moss region change with time. Consequently also the emission from moss varies.

The primary result from our simulations is that the area of the transition region can change substantially if there is a sizable change in the quasi-steady heating rate. For example, the area at 0.5 MK changes by about 40\% as the peak temperature in the flux tube varies between 1 and 4 MK. 
The plasma beta at the relevant heights is low enough that the magnetic field is minimally influenced by the enhanced gas pressure. The dominant effect of the pressure is to force the transition region downward to a place where the tube is more highly constricted. This alters the DEM($T$) curve, since the shape of the curve, including the ratio of coronal to transition region emission, depends on how the flux tube area varies with temperature. It may not be reasonable to independently specify a particular $A(T)$ dependence when modeling the solar atmosphere.
\cite{Warren_al_2010_b} found an $A(T)$ that reproduces the observed DEM($T$) of an active region core, but it is not obvious that this A(T) is consistent with the assumed steady heating rate.

\rev{Our model is simplified in several respects. Our 2D cylindrical description assumes a cylindrical symmetry around the loop central axis. Although there are some indications that this might not always be the case \cite{Malanushenko2013a}, we do not expect considerable differences of the results for moderate deviations of the shape of the cross-section. The structure and energy balance of the chromosphere are also simplified into an isothermal structure with no radiation transport, but our study focusses on layers at much higher temperature that should not be highly affected by this approximation. Finally, although our model implies a full MHD description, still it is not entirely self-consistent, since the heating is imposed artificially. A fully consistent model is out of the scope of this work, since we focus on the very specific question of the reaction of the low corona to heating.}

Another interesting result is the magnitude of the flux tube area expansion between the transition region and the corona. The area increases by approximately a factor of 2 to 3 over the first 3000 km above the chromosphere in our model, depending on the peak temperature. This is much less than in the models of \cite{Gabriel_1976,Athay1981a,Dowdy1986a,Warren_al_2010_b}.
Those models have a "throat" of rather extreme expansion, which we suggest may not be realistic. The Gabriel and Athay models use a potential field and assume that the flux tube is highly constricted at the {\it top} of the chromosphere. They further assume that the flux tube is isolated, with no surrounding field, which is probably a poor assumption, as discussed in \cite{Dahlburg_2005}.
Our \rev{"$\beta$-sensitive"} MHD model suggests that throats of extreme expansion are unlikely, although convective motions, not included here, may serve to enhance the effect. We note that the dependence of $A(T)$ on heating rate would be even stronger if such throats did exist.

\bigskip

\acknowledgements{We thank the anonymous referee for constructive comments and suggestions. We acknowledge support from Italian \emph{ Ministero dell'Universit\`a e Ricerca} and \emph{Agenzia Spaziale Italiana (ASI)},
 contract I/023/09/0 and I/015/07/0. PLUTO is developed at the Turin Astronomical Observatory in collaboration with the Department of Physics of the Turin University. We acknowledge the CINECA awards N. HP10CWS0PW and N. HP10B54VL7 under the ISCRA initiative, and the HPC facility (SCAN) of the INAF - Osservatorio Astronomico di Palermo, for the availability of high performance computing resources and support.
\emph{CHIANTI} is a collaborative project involving the \emph{NRL (USA)}, the \emph{Universities of Florence (Italy)} and
\emph{Cambridge (UK)}, and \emph{George Mason University (USA)}. The work of JAK was supported by the NASA Supporting Research and Technology Program.}
%
\bibliographystyle{apj}
\bibliography{biblio_max}

\end{document}